\documentclass[10,journal]{IEEEtran}
\usepackage{xcolor}
\usepackage{graphicx}
\usepackage[margin=0.75in]{geometry}

\usepackage{amsmath,amssymb,amsthm}
\usepackage{mathtools}
\usepackage{bm}

\usepackage{booktabs}
\usepackage{multirow}
\usepackage{siunitx}

\usepackage{enumitem}
\setlist{nosep}
\setitemize{leftmargin=*}
\usepackage{titlesec}

\usepackage{standalone}

\usepackage{tikz}
\usetikzlibrary{
  3d,
  arrows.meta,
  backgrounds,
  calc,
  positioning,
  shapes.geometric,
  shadows.blur
}
\usepackage{pgfplots}
\pgfplotsset{compat=1.18}
\usepgfplotslibrary{groupplots}

\usepackage[ruled,vlined,linesnumbered]{algorithm2e}
\SetKw{KwBreak}{break}

\definecolor{teal}{RGB}{0,128,128}
\definecolor{slate}{RGB}{112,128,144}
\definecolor{mainblue}{RGB}{25,70,140}
\definecolor{softgray}{RGB}{140,140,140}
\definecolor{aliceblue}{RGB}{240,248,255}

\newcommand{\bc}{\ensuremath{\bm{c}}}

\newcommand{\bg}{\ensuremath{\bm{g}}}

\newcommand{\bp}{\ensuremath{\bm{p}}}
\newcommand{\bq}{\ensuremath{\bm{q}}}
\newcommand{\br}{\ensuremath{\bm{r}}}
\newcommand{\bs}{\ensuremath{\bm{s}}}
\newcommand{\bt}{\ensuremath{\bm{t}}}
\newcommand{\bu}{\ensuremath{\bm{u}}}
\newcommand{\bv}{\ensuremath{\bm{v}}}

\newcommand{\bx}{\ensuremath{\bm{x}}}
\newcommand{\by}{\ensuremath{\bm{y}}}
\newcommand{\bz}{\ensuremath{\bm{z}}}

\newcommand{\bA}{\ensuremath{\bm{A}}}
\newcommand{\bB}{\ensuremath{\bm{B}}}
\newcommand{\bC}{\ensuremath{\bm{C}}}
\newcommand{\bD}{\ensuremath{\bm{D}}}
\newcommand{\bE}{\ensuremath{\bm{E}}}

\newcommand{\bG}{\ensuremath{\bm{G}}}

\newcommand{\bI}{\ensuremath{\bm{I}}}

\newcommand{\bP}{\ensuremath{\bm{P}}}
\newcommand{\bQ}{\ensuremath{\bm{Q}}}
\newcommand{\bR}{\ensuremath{\bm{R}}}

\newtheorem{assumption}{Assumption}

\newtheorem{theorem}{Theorem}
\newtheorem{corollary}{Corollary}

\newtheorem{problem}{Problem}
\usepackage{standalone}
\usepgfplotslibrary{fillbetween,groupplots}

\definecolor{darkblue}{RGB}{0, 0, 139}

\usepackage[colorlinks=true,
  linkcolor=mainblue,
  citecolor=mainblue,
  urlcolor=mainblue]{hyperref}
\newtagform{blue_eq}{\color{mainblue}(}{)}

\usetagform{blue_eq}
\usepackage{lettrine}
\usepackage{cite}
\usepackage{xcolor, colortbl}
\usepackage{graphicx}
\newcommand{\blueRule}{\arrayrulecolor{mainblue}\specialrule{1.2pt}{0.5pt}{0.5pt}\arrayrulecolor{black}}
\newcommand{\thinBlueRule}{\arrayrulecolor{mainblue!50}\specialrule{0.5pt}{0.5pt}{0.5pt}\arrayrulecolor{black}}

\makeatletter
\renewcommand{\section}{\@startsection{section}{1}{\z@}%
  {0.75ex plus 0.4ex minus 0.4ex}%
  {0.4ex plus 0.3ex minus 0ex}%
  {\normalfont\large \centering\scshape\color{black}}}

\renewcommand{\subsection}{\@startsection{subsection}{2}{\z@}%
  {0.75ex plus 0.4ex minus 0.4ex}%
  {0.3ex plus 0.2ex minus 0ex}%
  {\normalfont\normalsize\itshape\color{black}}}

\renewcommand{\subsubsection}{\@startsection{subsubsection}{3}{\z@}%
  {0.75ex plus 0.4ex minus 0.4ex}%
  {0.3ex plus 0.2ex minus 0ex}%
  {\normalfont\normalsize\itshape\color{mainblue}}}
\makeatother

\usepackage{makecell}

\newtheorem{lemma}{Lemma}

\usepackage[subtle]{savetrees}
\usepackage{array}
\newcolumntype{C}[1]{>{\centering\arraybackslash}m{#1}}

\usepackage{geometry}
\usepackage{etoolbox}
\usepackage{titlesec}

\title{\Large \color{mainblue}{\textsc{\textbf{Beyond Binary Controllability: \\ Reachability Margins for Power Systems via Scalable Screening}}}}

\author{Abdallah Alalem Albustami, Ahmad F. Tah$\text{a}^{\dagger}$, and Mohamad H. Kazma \thanks{{The authors are all affiliated with the Civil and Environmental Engineering and Electrical and Computer Engineering Departments at Vanderbilt University, Nashville, Tennessee. This work is supported by the National Science Foundation under Grant 2152450. $^\dagger$Corresponding author. Email: \url{ahmad.taha@vanderbilt.edu.}}}\vspace{-0.6cm}}
\date{\today}

\begin{document}
	
	\maketitle
	\thispagestyle{empty}
	\pagestyle{empty}
	
	\begin{abstract}
		This paper studies unreachability screening for power systems, focusing on the inadequate control authority that arises under stressed grid conditions. Its main result is an infeasibility \textit{margin} for linearized differential algebraic equation models that quantifies how far a prescribed terminal target lies from the reachable set in finite time. The paper specifically develops a conic benchmark method together with two first order methods based on projected subgradients and bundle updates for scalably repeated screening. Studies on IEEE benchmark networks show reduced terminal reachability under stressed conditions and tighter actuation limits, while first order methods remain accurate and substantially faster on large systems. Comparisons with standard reachability methods such as CORA further demonstrate the usefulness and scalability of the proposed infeasibility margins. All results and codes are included in a Github repository. 
	\end{abstract}
	
	
	\section{Introduction and Paper Contributions}
	\label{sec:intro}
	
	\lettrine{P}{OWER} systems operate around economic dispatch and optimal power flow (OPF) setpoints. Their dynamics are studied locally through linearized versions of the nonlinear differential-algebraic equation (DAE) models.  For a multi-machine power network, DAEs couple dynamic states such as frequencies with algebraic, network power flow constraints and variables (such as voltages and angles) \cite{Kundur1994,SauerPai1997,Milano2010}. Much of the literature removes the exact dynamic-algebraic coupling to obtain ODE models, simplifying analysis and enabling known control methods primarily developed for ODEs.
	
	\vspace{0.15cm}
	
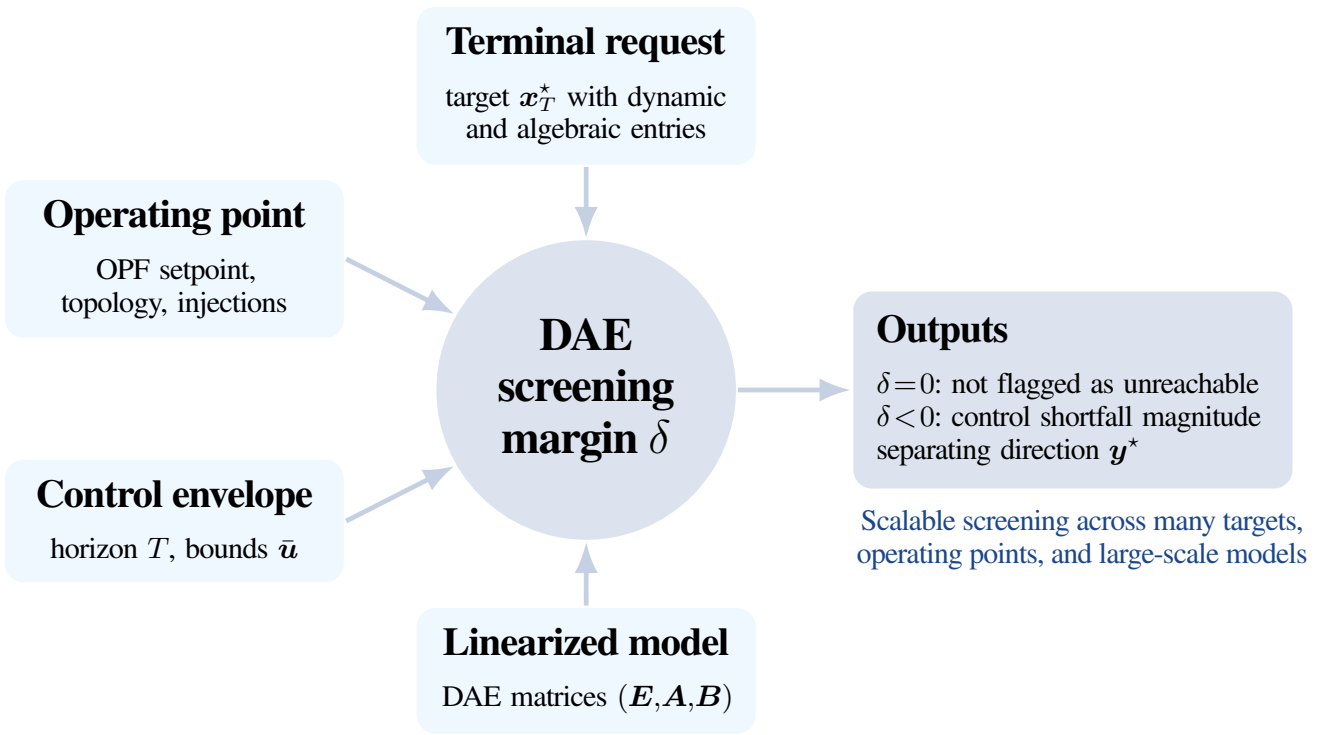
\begin{figure}[t]
		\centering
		\resizebox{\columnwidth}{!}{%
			\begin{tikzpicture}[
				>=Latex,
				font=\small,
				line width=1.45pt,
				inputbox/.style={
					rounded corners=2mm,
					fill=aliceblue,
					align=center,
					inner sep=6pt,
					text width=3.2cm},
				core/.style={
					very thick,
					circle,
					fill=mainblue!15,
					minimum size=3.2cm,
					align=center},
				outputbox/.style={
					rounded corners=2mm,
					fill=mainblue!15,
					align=left,
					inner sep=7pt,
					text width=4.2cm}]
				\node[inputbox] (opf) at (-4.4, 1.4) {\textbf{\large Operating point}\\ [2mm] OPF setpoint, topology, injections};
				\node[inputbox] (target) at (0, 3.25) {\textbf{\large Terminal request}\\ [2mm] target \(\bx_T^\star\) with dynamic and algebraic entries};
				\node[inputbox] (limits) at (-4.4,-1.4) {\textbf{\large Control envelope}\\ [2mm] horizon \(T\), bounds \(\bar{\bu}\)};
				\node[inputbox] (model) at (0,-3) {\textbf{\large Linearized model}\\ [2mm] DAE matrices \((\bE,\bA,\bB)\)};
				
				\node[core] (screen) at (0,0) {\Large \textbf{DAE}\\ [2mm] \Large\textbf{screening}\\[1mm] \Large \textbf{margin} \(\delta\)};
				
				\node[outputbox] (out) at (5.2,0) {\textbf{\large Outputs}\\ [2mm]
					\(\delta=0\): not flagged as unreachable\\
					\(\delta<0\): control shortfall magnitude\\
					separating direction \(\by^\star\)};
				
				\draw[->,mainblue!25] (opf.east) -- (screen.150);
				\draw[->,mainblue!25] (limits.east) -- (screen.210);
				\draw[->,mainblue!25] (target.south) -- (screen.90);
				\draw[->,mainblue!25] (model.north) -- (screen.270);
				\draw[->,mainblue!25] (screen.east) -- (out.west);
				
				\node[align=center,mainblue] at (5.3,-1.6) {Scalable screening across  many targets, \\ operating points, and large-scale models};
		\end{tikzpicture}}
        \vspace{-0.5cm}
		\caption{Inputs and outputs of the proposed screening tool. }
		\label{fig:cdc_screening_schematic}
	\end{figure}

	\textsc{\textbf{Paper's objective and scope}.} The paper's objective is to study this screening problem:  given an OPF setpoint, a schedule of load and renewable injections, a linearized DAE model, a control horizon of $T$ seconds, bounds on the control input deviations, and an arbitrary terminal target $\bx_T^\star$, a prescribed value of the deviation $\Delta \bx(T)$, compute a scalable margin $\delta\le 0$ that quantifies the difficulty of this engineering problem. A negative $\delta$ certifies that the target is unreachable and equals the negative distance to the terminal reachable set, while $\delta=0$ means the target is not flagged. The idea is presented in Fig.~\ref{fig:cdc_screening_schematic}.
	
	For multi-machine power network models, the Kalman controllability test is numerically unreliable even when the PBH test certifies controllability \cite{paige1981properties}, and models controllable in the rank sense remain practically uncontrollable along weak directions where the required control effort is prohibitive \cite{pasqualetti2014controllability}. In this context, absolute controllability is not very meaningful, as some state targets are not interesting from the power system side and hence their reachability is not relevant. However, this investigation becomes pertinent in renewables-rich operation, because the operating point, the local Jacobians, and the effective control authority all vary with dispatch and network stress. With more renewables, synchronous machines and their controls are displaced, which reduces the available control authority and typically shrinks the reachable space under bounded inputs \cite{milano2018foundations}.
	
	For the problem we study in this paper, some terminal state requests may remain easy to realize, while others become difficult because the horizon $T$ can be short, the controller bounds are tight, or the algebraic network closure prevents the full terminal state from matching the request. A useful screening tool therefore needs to depend explicitly on the inputs described in Fig.~\ref{fig:cdc_screening_schematic}. 
	
	The central modeling point is that the terminal target lives on the \emph{full} DAE-consistent manifold. Reduced ODE models remain useful, but they hide failure mechanisms carried by algebraic variables such as voltages and network-balance quantities. In contrast, the target in this paper may simultaneously request acceptable recovery of frequencies, internal EMF variables, and other dynamic states. The resulting tool herein screens frequency control and voltage control reachability in one scalable computation. The relevant literature is reviewed next. 
	
	\vspace{0.15cm}
	
	\textsc{\textbf{Brief literature review}.} The relevant literature spans DAE system theory, scalable reachability computation, and power system applications. The studies \cite{MalabreKuceraZagalak1990,Ozcadiran1989} characterize reachable and controllable subspaces through matrix-pencil structure and DAE indices. The authors in this extensive survey \cite{BergerReis2013} systematize the different controllability notions that arise once consistency constraints are present, and the study \cite{FengLam2015} develops reachable set estimation for singular systems with explicit treatment of admissible initial conditions. The study \cite{TranJohnson2019} treats high-index linear DAE reachability by separating slow and algebraic dynamics through consistency enforcing constructions. These contributions explain how DAEs differ from ODEs and why algebraic constraints must be handled explicitly, but they do not formulate a target-specific screening metric.
	
	In the scalable reachability analysis more broadly, the study \cite{girard2005reachability} propagates uncertain linear reachable sets with zonotopes (convex polytopes). The authors in \cite{leguernicgirard2010nahs} develop support function propagation for linear systems by repeatedly maximizing linear functionals rather than enumerating full sets. The work \cite{althoff2010reachability} uses conservative linearization to bound nonlinear reachable sets, and the study \cite{althofffrehse2016cdc} combines zonotopes to improve computational scalability for large systems. These works make clear why support functions are important: they encode the directional action of a high-dimensional reachable set, support separating-hyperplane arguments, and avoid explicit construction of a complicated set in every computation. Our work here, instead of propagating an entire reachable tube, uses the support function of the terminal reachable set to compute an exact \textit{target-to-set} \textit{shortfall} for a prescribed terminal request.
	
	In power systems, reachable set ideas mainly support transient stability assessment and formal verification. The study \cite{WadduwageAnnakkageChaudhuri2013} investigates transient stability assessment through reachable set computations. The study \cite{AlthoffKrogh2014DAE} investigates reachability analysis for nonlinear differential-algebraic systems and shows how set-propagation ideas adapt to DAE structure. The dissertation \cite{Althoff2014Power} develops formal and compositional reachability analysis for power systems under uncertainty, including decomposition strategies for scalability, conservative set propagation around operating points, and reachable set based verification. Another dissertation \cite{Elguindy2018Reachability} extends this work and develops transient stability assessment, region of attraction estimation, LPV-based approximations, and reachability-assisted control design for large power networks; see also studies \cite{ElGuindy2017Compositional,ElGuindy2017LPV} that delineate the contributions of the dissertation~\cite{Elguindy2018Reachability}. These studies compute (or over-approximate) reachable sets to certify safety, stability, or control performance. They do \textit{not} derive an exact Euclidean distance from a prescribed full terminal target to the terminal reachable set of a linearized power system DAE model. Our recent work~\cite{11552847} paper develops a scalable, target-aware reinterpretation of the PBH controllability test by using dual infeasibility vectors to \textit{certify} unreachable terminal states. This study extends our work with a focus on DAEs, power systems, and continuous-time (CT) dynamic models compared to the discrete-time (DT) ODEs in~\cite{11552847}.
	
	\vspace{0.15cm}
	
	\textsc{\textbf{Paper contributions. }} The paper contributions are:
	\begin{itemize}
		\item We formulate terminal reachability directly on the DAE-consistent manifold, so the target may include both differential and algebraic power system variables.
		\item For index-1 linearized power system DAEs with bounded controls, we derive an explicit support function representation of the full terminal reachable set and a screening margin $\delta$ that equals the negative Euclidean distance to that set when the target is infeasible. This quantifies the difficulty in steering the prescribed system to the target. 
		\item We present three computational routes tailored to screening: a quadrature-based conic approximation, a projected subgradient method, and a bundle-type method. The methods are endowed with different properties and scale nicely. Thorough case studies are included to corroborate the theory. 
	\end{itemize}
	Section~\ref{sec:problem} introduces the model and the screening problem. Section~\ref{sec:theory} derives the margin and its geometric meaning. Section~\ref{sec:compute} discusses how to compute the metric efficiently. Section~\ref{sec:casestudies} presents case studies and Section~\ref{sec:conclusion} concludes the paper.
	

	\section{System Model and Problem Formulation}
	\label{sec:problem}
	
	We start from the nonlinear DAE model~\cite{Kundur1994,bazrafshan2019coupling}
	\begin{subequations}\label{eq:nl_dae_compact}
		\begin{align}
			\textit{generator physics:}\qquad \dot{\bx}_d(t) &= \mathbf{f}_d\big(\bx_d(t),\bx_a(t),\bu(t)\big), \label{eq:nl_dae_compact_a}\\
			\textit{network power flow:}\qquad \mathbf{0} &= \mathbf{f}_a\big(\bx_d(t),\bx_a(t),\bu(t)\big), \label{eq:nl_dae_compact_b}
		\end{align}
	\end{subequations}
	where \(\bx_d(t)\in\mathbb{R}^{n_d}\) collects the differential states, \(\bx_a(t)\in\mathbb{R}^{n_a}\) collects the algebraic states, and \(\bu(t)\in\mathbb{R}^{m}\) is the control input. Around an equilibrium \((\bar{\bx}_d,\bar{\bx}_a,\bar{\bu})\), the linearized model is
	\begin{subequations}\label{eq:lin_dae_split_relabel}
		\begin{align}
			\dot{\bx}_d(t)
			&=
			\bA_{dd}\bx_d(t)+\bA_{da}\bx_a(t)+\bB_u\bu(t), \label{eq:lin_dae_split_relabel_a}\\
			\mathbf{0}
			&=
			\bA_{ad}\bx_d(t)+\bA_{aa}\bx_a(t)+\bB_a\bu(t). \label{eq:lin_dae_split_relabel_b}
		\end{align}
	\end{subequations}
    or
	\begin{equation}\label{eq:descriptor_form}
		\bE\dot{\bx}(t)=\bA\bx(t)+\bB\bu(t),
	\end{equation}
	with
	\[
	\bE=
	\begin{bmatrix}
		\bI_{n_d} & \mathbf{0}\\
		\mathbf{0} & \mathbf{0}
	\end{bmatrix},
	\quad
	\bA=
	\begin{bmatrix}
		\bA_{dd} & \bA_{da}\\
		\bA_{ad} & \bA_{aa}
	\end{bmatrix},
	\quad
	\bB=
	\begin{bmatrix}
		\bB_u\\
		\bB_a
	\end{bmatrix}.
	\] \vspace{0.1cm}
	
	\begin{table}[t]
		\caption{Power system meaning of the main vectors in the linearized screening model.}
		\label{tab:state_summary}
		\centering
		\scriptsize
		\renewcommand{\arraystretch}{1.15}
		\begin{tabular}{p{0.6cm}p{4cm}p{2.8cm}}
			\blueRule
			\rowcolor{mainblue!10}
			\textbf{Symbol} & \textbf{Typical contents} & \textbf{Role in screening} \\
			\thinBlueRule
			\(\bx_d\) & rotor angles, frequencies, internal EMFs, governor states, inverter-controller states & dynamic quantities that evolve over \([0,T]\) \\
			\thinBlueRule
			\(\bx_a\) & bus voltages, phase angles, nodal balance variables, algebraic network quantities & algebraic quantities that must close consistently at the terminal time \\
			\thinBlueRule
			\(\bu\) & AGC/secondary-control actions, inverter setpoints, FACTS/HVDC modulation, controllable load actions & bounded actuation used to move the terminal state \\
			\thinBlueRule
			\(\bx_T^\star\) & selected terminal values from both \(\bx_d\) and \(\bx_a\) & screening target to be tested for reachability \\
			\thinBlueRule
			\(\boldsymbol{\xi}_T\) & target after removing the free response & forced displacement that the controls must generate \\
			\blueRule
		\end{tabular}
	\end{table}
	
	Tab.~\ref{tab:state_summary} summarizes the main variables in the power system model and the screening tool. We note the following: \textit{(i)} in many classical models \(\bB_a=\mathbf{0}\) because generators' control inputs do not appear in the algebraic power flow constraints, but we retain \(\bB_a\) because several grid control strategies (i.e., demand response or renewables curtailment) act directly on the algebraic constraints. \textit{(ii)} Likewise, loads and renewable injections do not appear as separate variables in~\eqref{eq:lin_dae_split_relabel} as their influence is absorbed into the operating point and the Jacobian matrices. There is no loss of generality here, as the model can still accommodate their deviations. The math is simpler without them, and hence they are omitted. 
	
	\begin{assumption}[Index-1 regularity]
		\label{ass:index1}
		Matrix \(\bA_{aa}\) is nonsingular, and the descriptor pair \((\bE,\bA)\) is regular and impulse-free.
	\end{assumption}
	
	Under Assumption~\ref{ass:index1}, which is common in the power systems literature and is satisfied for a host of standard benchmark models near regular operating points \cite{Milano2010, GrossTrennWirsen2016}, the algebraic state is given by
	\begin{equation}\label{eq:alg_elim}
		\bx_a=\bC_r\bx_d+\bD_r\bu, \; \bC_r:=-\bA_{aa}^{-1}\bA_{ad}, \; \bD_r:=-\bA_{aa}^{-1}\bB_a.
	\end{equation}
	 Substituting into~\eqref{eq:lin_dae_split_relabel_a} yields the reduced dynamics
	\begin{equation}\label{eq:reduced_ode}
		\dot{\bx}_d(t)=\bA_r\bx_d(t)+\bB_r\bu(t),
	\end{equation}
	with \(\bA_r:=\bA_{dd}-\bA_{da}\bA_{aa}^{-1}\bA_{ad}\) and \(\bB_r:=\bB_u-\bA_{da}\bA_{aa}^{-1}\bB_a\). Matrix $\bC_r$ is the linearized algebraic-recovery map: it tells us how a perturbation in the differential state instantaneously propagates to the algebraic variables through the power flow equations. 
	
	Given symmetric controller limits \(\bar{\bu}\in\mathbb{R}^m_{>0}\) on the deviations of the generators' control actions $\bu(t)$ (recall that $\bu(t)$ is the deviation in the setpoints), we use the admissible input class
	\[
	\mathcal{U}_T
	:=
	\left\{
	\bu\in L_\infty([0,T],\mathbb{R}^m)\ \middle|\ |\bu(t)|\le \bar{\bu}\ \text{a.e. on }[0,T]
	\right\}.
	\]
	By \eqref{eq:alg_elim}, the algebraic closure holds for all \(t\); at the terminal time it is determined by the admissible endpoint input value. We therefore introduce an explicit terminal control variable \(\bu_T\in\mathbb{R}^m\) satisfying \(|\bu_T|\le \bar{\bu}\). This avoids committing the theory to a specific endpoint convention. If the implementation is continuous at \(T\), one identifies \(\bu_T=\bu(T)\); if a sampled implementation uses a pre-jump endpoint convention, one identifies \(\bu_T=\bu(T^-)\). The analysis below only needs the admissible endpoint value \(\bu_T\). For any admissible endpoint control value \(\bu_T\), define
	\begin{equation*}
		\small \begin{aligned}
			\mathcal{M}_T(\bu_T)
			:=
			\Biggl\{
			\begin{bmatrix}
				\bx_d\\
				\bx_a
			\end{bmatrix}
			\in\mathbb{R}^{n_d+n_a}
			\;\Biggm|\;
			\bA_{ad}\bx_d+\bA_{aa}\bx_a+\bB_a\bu_T=\mathbf{0}
			\Biggr\}.
		\end{aligned}
	\end{equation*}
	We also use the admissible terminal manifold $\mathcal{M}_T
	:=
	\bigcup_{|\bu_T|\le \bar{\bu}} \mathcal{M}_T(\bu_T).$
	Let $\bx_{d,0}$ be a consistent initial differential state and let
	$\bx_T^\star = [
	\bx_{d,T}^{\star,\top}, \; 
	\bx_{a,T}^{\star,\top}]^{\top} $
	be the desired terminal state, or a terminal state value whose (un)reachability we seek to quantify. The free terminal response is defined as $\bx_T^{\rm fr}$ and the forced terminal operator induced by \((\bu(\cdot),\bu_T)\) is defined as $\mathbf{\Phi}_T(\bu,\bu_T)$. We also define \(\bP:=\bigl[\begin{smallmatrix}\bI\\ \bC_r\end{smallmatrix}\bigr]\) and \(\bG:=\bigl[\begin{smallmatrix}\mathbf{0}\\ \bD_r\end{smallmatrix}\bigr]\). We can now write
	\begin{equation*}
		\begin{aligned}
			\bx(T)&=\bx_T^{\rm fr}+\mathbf{\Phi}_T(\bu,\bu_T), \quad \bx_T^{\rm fr}:=\bP e^{\bA_rT}\bx_{d,0},\\
			\mathbf{\Phi}_T(\bu,\bu_T)&:=\bP\int_0^T e^{\bA_r(T-\tau)}\bB_r\bu(\tau)\,d\tau+\bG\bu_T.
		\end{aligned}
	\end{equation*}
	Accordingly,
	\[
	\mathcal{R}_T^{\rm full}
	:=
	\left\{
	\bx_T^{\rm fr}+\mathbf{\Phi}_T(\bu,\bu_T)\ \middle|\ \bu\in\mathcal{U}_T,\ |\bu_T|\le \bar{\bu}
	\right\},
	\]
	\[
	\mathcal{R}_T^{\rm forc}
	:=
	\left\{
	\mathbf{\Phi}_T(\bu,\bu_T)\ \middle|\ \bu\in\mathcal{U}_T,\ |\bu_T|\le \bar{\bu}
	\right\}.
	\]
	By construction, \(\mathcal{R}_T^{\rm full}\subseteq \mathcal{M}_T\). Therefore, any target outside \(\mathcal{M}_T\) is immediately impossible because it violates terminal algebraic consistency for every admissible endpoint control.
	
	We assume throughout that \(\mathcal{R}_T^{\rm forc}\) is nonempty, compact, and convex. For a desired terminal target \(\bx_T^\star\), let \(\boldsymbol{\xi}_T:=\bx_T^\star-\bx_T^{\rm fr}\) define the target after removing the free response. Then \(\bx_T^\star\in\mathcal{R}_T^{\rm full}\) if and only if \(\boldsymbol{\xi}_T\in\mathcal{R}_T^{\rm forc}\). The problem formulaiton is given next. 
	
	\begin{problem}\label{prob:screening}
		Given an OPF operating point, a horizon of \(T\) seconds, admissible control limits \(\bar{\bu}\), the linearized state-space matrices \((\bE,\bA,\bB)\) in~\eqref{eq:descriptor_form}, and a desired full terminal target \(\bx_T^\star\), quantify whether the target is unreachable and, if so, compute how far it is from the terminal reachable set via margin $\delta$. 
	\end{problem}
	
	\vspace{-0.2cm}
	\section{Margin Formulation on DAE Manifolds}
	\label{sec:theory}
	In this section, we present the simple theoretical result that allows us to compute the margin $\delta$. To do so, we need to define support functions. Readers familiar with support functions can skip the next paragraph. 
	
	For a nonempty set \(\mathcal{S}\subset\mathbb{R}^n\), the support function \(h_{\mathcal{S}}(\by):=\sup_{\bx\in\mathcal{S}}\by^\top\bx\) gives the maximum projection of the set along the direction \(\by\). The support function is a convex function that maps a vector to the maximum inner product over that set, essentially measuring the distance of the set's supporting hyperplane from the origin. For example, if \(\by\) places weights on frequencies, internal EMF states, voltages, phase angles, and other terminal coordinates, then \(h_{\mathcal{R}_T^{\rm forc}}(\by)\) is the largest terminal displacement that the admissible controls can produce along that weighted combination of quantities. 
	
	This interpretation is standard in reachability analysis, where support functions summarize high-dimensional reachable sets through linear functionals, support separating hyperplane arguments, and enable scalable outer approximations without explicit set gridding \cite{leguernicgirard2010nahs,althofffrehse2016cdc}. 
	This same idea makes the screening problem meaningful and scalable. Rather than constructing the full terminal reachable set, the paper works with its directional action and recovers an \textbf{\textit{exact target-to-set shortfall}} from that information.
	
				
				
				
				
				
	
	Having defined support functions and introduced the model and the terminal manifold, the next step is to characterize this directional action explicitly for the index-1 power system DAE. The following lemma provides this result. 
	
	\begin{lemma}
		\label{lem:support}
		Let Assumption~\ref{ass:index1} hold, and assume \(\mathcal{R}_T^{\rm forc}\) is nonempty, compact, and convex. Then \(\mathcal{R}_T^{\rm full}\subseteq \mathcal{M}_T\), and for every \(\by=[\by_d^\top\ \by_a^\top]^\top\in\mathbb{R}^{n_d+n_a}\),
		\begin{equation}\label{eq:full_support_formula}
			\begin{aligned}
				h_{\mathcal{R}_T^{\rm forc}}(\by)
				=\;&
				\int_0^T
				\left\|
				\bR_{\bar u}\,\bB_r^\top e^{\bA_r^\top(T-\tau)}\bP^\top\by
				\right\|_1\,d\tau\\
				&+
				\left\|
				\bR_{\bar u}\,\bD_r^\top\by_a
				\right\|_1,
			\end{aligned}
		\end{equation}
		where \(\bR_{\bar u}:=\mathrm{diag}(\bar{\bu})\). Consequently,
		\begin{equation}\label{eq:full_support_shifted}
			h_{\mathcal{R}_T^{\rm full}}(\by)
			=
			\by^\top\bx_T^{\rm fr}+h_{\mathcal{R}_T^{\rm forc}}(\by).
		\end{equation}
	\end{lemma}
	
	The proof is given in Appendix~\ref{app:proof_lem}.
	Targets outside \(\mathcal{M}_T\) can be screened out immediately. The next result handles the nontrivial case \(\bx_T^\star\in\mathcal{M}_T\), where algebraic consistency is possible but reachability is not guaranteed. This result addresses Problem~\ref{prob:screening}.
	\begin{theorem}
		\label{thm:margin}
		Let \(\bx_T^\star\in\mathcal{M}_T\). Define
		\begin{equation}\label{eq:ct_mcu_def}
			\delta(\bx_T^\star)
			:=
			\min_{\|\by\|_2\le 1}
			\left(
			h_{\mathcal{R}_T^{\rm forc}}(\by)-\by^\top\boldsymbol{\xi}_T
			\right),
		\end{equation}
		then:
		\vspace{-0.3cm}
		\begin{equation}\label{eq:screening_identity}
			\delta(\bx_T^\star)
			=
			\begin{cases}
				0, & \bx_T^\star\in\mathcal{R}_T^{\rm full},\\[1mm]
				-\mathrm{dist}\big(\bx_T^\star,\mathcal{R}_T^{\rm full}\big), & \bx_T^\star\notin\mathcal{R}_T^{\rm full},
			\end{cases}
		\end{equation}
		and equivalently,
		\begin{equation}\label{eq:dual_distance_formula}
			\delta(\bx_T^\star)
			= -
			\max_{\|\by\|_2\le 1}
			\left(
			\by^\top\boldsymbol{\xi}_T-h_{\mathcal{R}_T^{\rm forc}}(\by)
			\right).
		\end{equation}
		If \(\bx_T^\star\notin\mathcal{R}_T^{\rm full}\), any nonzero maximizer in~\eqref{eq:dual_distance_formula} defines a separating direction for the unreachable target.
	\end{theorem}
	
	The proof is given in Appendix~\ref{app:proof_thm}.
	Theorem~\ref{thm:margin} shows that the screening margin is a one-sided control shortfall metric. It returns zero when the target is not certified as unreachable and becomes negative precisely when the target lies outside the full terminal reachable set. The direction \(\by^\star\) indicates where the terminal request most strongly exceeds the attainable directional capacity. In power system terms, the dominant entries of \(\by^\star\) localize the shortfall at specific machines (frequency entries) or buses (voltage and angle entries), and \(\bx_T^{\rm fr}+\boldsymbol{\xi}_T-|\delta|\,\by^\star\) is the nearest attainable terminal state, hence a minimally relaxed target. Typical requests include full or fractional recovery of a disturbance and redispatch toward a neighboring equilibrium, with \(T\) set by the secondary-control window and \(\bar{\bu}\) by governor and inverter headroom.
	\begin{corollary}
		\label{cor:monotonicity}
		Consider two screening instances defined on the same terminal state space, with full terminal reachable sets \(\mathcal{R}_{T,1}^{\rm full}\) and \(\mathcal{R}_{T,2}^{\rm full}\). Let \(\bar{\bx}_T\) denote a fixed terminal request screened against both instances. If $\mathcal{R}_{T,1}^{\rm full}\subseteq \mathcal{R}_{T,2}^{\rm full},$
		then $
		\delta_1(\bar{\bx}_T)\le \delta_2(\bar{\bx}_T)\le 0.$
		In particular, if \(\bar{\bx}_T\) is unreachable for both instances, then \(|\delta_1(\bar{\bx}_T)|\ge |\delta_2(\bar{\bx}_T)|\).
	\end{corollary}
	The proof is given in Appendix~\ref{app:proof_cor}.
	Corollary~\ref{cor:monotonicity} formalizes the sense in which the metric gets worse for a \textit{more unreachable} instance, i.e., if a design change shrinks the terminal reachable set, then the margin cannot improve for the same terminal request. 

    \vspace{-0.05cm}
	\section{Computing the Margin}
	\label{sec:compute}
	The simple results in the previous section reduce screening to repeated evaluation of the support function in Lemma~\ref{lem:support} and the optimization problem in Theorem~\ref{thm:margin}. In this section, we develop three approaches to solve the optimization. Approach I approximates the \textit{exact} margin by quadrature rules and then solves a finite-dimensional convex conic problem. Approach II keeps the CT model intact and applies projected subgradient ascent to the dual objective. Approach III uses the same CT dual formulation but stabilizes the search with a bundle-type model. 

    \vspace{-0.05cm}
	\subsection{Approach I: Quadrature and Conic Optimization}
	Approach I targets a high-fidelity approximation of~\eqref{eq:ct_mcu_def}. The key observation is that the support function in Lemma~\ref{lem:support} separates into a CT dynamic integral and a finite-dimensional terminal algebraic term. We therefore approximate only the integral by Gauss-Legendre quadrature \cite{dahlquist2003numerical}. Let \(\{\hat\zeta_j,\hat w_j\}_{j=1}^{N_q}\) be the Gauss-Legendre nodes and weights on \([-1,1]\), the roots of the degree-\(N_q\) Legendre polynomial and their quadrature weights, and define, for \(j=1,\dots,N_q\), the following terms
	\[
	\zeta_j=\frac{T}{2}(\hat\zeta_j+1),
	\quad
	w_j=\frac{T}{2}\hat w_j, \quad \bQ_j:=\bR_{\bar u}\,\bB_r^\top e^{\bA_r^\top(T-\zeta_j)}\bP^\top.
	\]
	Then
	\[
	\int_0^T
	\left\|
	\bR_{\bar u}\,\bB_r^\top
	e^{\bA_r^\top(T-\tau)}\bP^\top \by
	\right\|_1 d\tau
	\approx
	\sum_{j=1}^{N_q}
	w_j
	\left\|
	\bQ_j \by
	\right\|_1,
	\]
	and the margin is approximated by
	\begin{equation}\label{eq:quad_margin}
		\small \hspace{-0.5cm}\delta^{(q)}(\bx_T^\star)
		=
		\min_{\|\by\|_2\le 1}
		\left[
		\sum_{j=1}^{N_q} w_j\|\bQ_j\by\|_1
		+
		\|\bR_{\bar u}\bD_r^\top\by_a\|_1
		-\by^\top\boldsymbol{\xi}_T
		\right].
	\end{equation}
	Introducing slack variables \(\bt_j\in\mathbb{R}^m\) and \(\bs\in\mathbb{R}^m\) yields:
	\begin{equation}\label{eq:quad_socp}
		\begin{aligned}
			\delta^{(q)}(\bx_T^\star)=
			\min_{\by,\{t_j\},s}\quad
			&
			\sum_{j=1}^{N_q} w_j \mathbf{1}^\top \bt_j
			+
			\mathbf{1}^\top \bs
			-
			\by^\top \boldsymbol\xi_T
			\\
			\text{subject to}\quad
			&
			-\bt_j \le \bQ_j \by \le \bt_j,
			\qquad j=1,\dots,N_q,
			\\
			&
			-\bs \le \bR_{\bar \bu}\bD_r^\top \by_a \le \bs,
			\\
			&
			\|\by\|_2\le 1,
			\quad
			\bt_j\ge \textbf{0},
			\quad
			\bs\ge \textbf{0}.
		\end{aligned}
	\end{equation}
    
	This is a finite-dimensional second order cone program (SOCP) and may therefore be solved globally by standard conic optimization solvers. It is always feasible since \(\by=\mathbf{0}\), \(\bt_j=\mathbf{0}\), and \(\bs=\mathbf{0}\) satisfy all constraints and its objective is bounded below by \(-\|\boldsymbol{\xi}_T\|_2\), so a minimizer exists.
    Its main cost comes from repeatedly evaluating the matrix exponentials embedded in the quadrature matrices \(\bQ_j\) and from the number \(N_q\) of quadrature nodes. This formulation can become expensive if many operating points or many target requests must be screened. 

	\subsection{Approach II: Projected Subgradient Method}
	The first approach, while \textit{benchmark worthy}, can be computationally challenging for large-scale power networks. To that end, the second approach works directly with the CT dual and never discretizes the DAE model at the formulation level. Define
	\begin{equation}\label{eq:J_of_y}
		\begin{aligned}
			J(\by)
			:=\;&
			\by^\top\boldsymbol{\xi}_T -\int_0^T
			\left\|
			\bR_{\bar u}\,\bB_r^\top e^{\bA_r^\top(T-\tau)}\bP^\top\by
			\right\|_1\,d\tau\\
			&-
			\left\|
			\bR_{\bar u}\,\bD_r^\top\by_a
			\right\|_1.
		\end{aligned}
	\end{equation}
	Then Theorem~\ref{thm:margin} implies
	\(
	-\delta(\bx_T^\star)=\max_{\|\by\|_2\le 1}J(\by).
	\)
	The objective \(J(\by)\) is concave but generally nonsmooth because of the \(\ell_1\)-norm terms. Consequently, one natural computational route is a projected subgradient method, in which a subgradient ascent step is taken on \(J(\by)\) and the iterate is then projected back onto the feasible set \(\{\by\in\mathbb{R}^{n_d+n_a}\mid \|\by\|_2\le 1\}\); see, e.g., \cite{Boyd2008Subgradient}. A second option is to use a bundle-type method~\cite{kiwiel2006methods}, which retains several past subgradients and function values to construct a stabilized piecewise-linear model of the nonsmooth objective and often yields more reliable practical progress than plain subgradient steps; see \cite{kiwiel2006methods}. 
	This section focuses on a projected subgradient method. This route keeps the DAE model continuous in time and moves all optimization to the finite-dimensional variable \(\by\). For a given \(\by\), let \(\bv(\tau)\) solve the adjoint reduced dynamics
	\[
	-\dot{\bv}(\tau)=\bA_r^\top\bv(\tau),
	\qquad
	\bv(T)=\bP^\top\by.
	\]
	Then \(\bv(\tau)=e^{\bA_r^\top(T-\tau)}\bP^\top\by\), and a valid extreme control is
	\[
	\bu_{\by}(\tau)=\bR_{\bar u}\,\mathrm{sign}\!\big(\bB_r^\top\bv(\tau)\big),
	\qquad
	\bu_{T,\by}=\bR_{\bar u}\,\mathrm{sign}\!\big(\bD_r^\top\by_a\big),
	\]
	with any subgradient choice at zero. The extreme terminal point is
	\[
	\mathbf{\Phi}_T(\bu_{\by},\bu_{T,\by})
	=
	\bP\int_0^T e^{\bA_r(T-\tau)}\bB_r\bu_{\by}(\tau)\,d\tau
	+\bG\bu_{T,\by},
	\]
	so a valid subgradient of \(J(\by)\) is
	\begin{equation}\label{eq:subgradient}
		\bg(\by)=\boldsymbol{\xi}_T-\mathbf{\Phi}_T(\bu_{\by},\bu_{T,\by}).
	\end{equation}
	In the numerical implementation, both \(J(\by)\) and \(\bg(\by)\) are evaluated with \(N_q\) quadrature nodes, so the projected subgradient routine is indexed by the triplet \((N_q,K,\alpha_k)\), where \(K\) is the iteration budget.
	
	A projected subgradient ascent step is therefore
	\[
	\by^{k+1}
	=
	\Pi_{\|\cdot\|_2\le 1}\big(\by^k+\alpha_k\bg(\by^k)\big),
	\]
	where \(\Pi_{\|\cdot\|_2\le 1}\) denotes Euclidean projection onto the unit ball and \(\alpha_k>0\) is a stepsize \cite{Boyd2008Subgradient}. Standard diminishing stepsize rules such as \(\alpha_k=\alpha_0/\sqrt{k+1}\) keep the iteration stable in the nonsmooth setting. If \(\{\by^k\}_{k=0}^K\) denotes the generated iterates, a practical estimate of the screening margin is $
	\widehat{\delta}_{\rm sg}^{(K)}(\bx_T^\star)
	:=
	-\max_{0\le k\le K}J(\by^k).$
	Whenever \(J(\by^k)>0\) for some iterate, the corresponding estimate is strictly negative and terminal infeasibility is certified.
	
	\subsection{Approach III: Bundle-Type Method}
	
	While projected subgradient ascent uses only the most recent subgradient information, a bundle-type method~\cite{kiwiel2006methods} retains several past function values and subgradients in order to build a local piecewise-linear model of the concave nonsmooth objective \(J(\by)\). This is often more stable in practice, especially when the dual landscape contains many changes in active signs induced by the \(\ell_1\)-norm terms. The key idea is to replace the true objective locally by the best available collection of affine upper models and then maximize this model with an explicit proximal regularization over the Euclidean unit ball.

	Suppose that at bundle points \(\by^{(i)}\), \(i=1,\dots,p\), we have already computed objective values \(J(\by^{(i)})\) and valid subgradients \(\bg^{(i)}\in\partial J(\by^{(i)})\). Since \(J(\by)\) is concave, every subgradient gives the global affine upper model
	\[
	J(\by)\le J(\by^{(i)})+{\bg^{(i)}}^\top(\by-\by^{(i)}),\qquad i=1,\dots,p.
	\]
	The resulting piecewise-linear model is
	\begin{equation}\label{eq:bundle_model_root_cdc}
		\widehat{J}_p(\by)
		:=
		\min_{1\le i\le p}
		\left(
		J(\by^{(i)})+{\bg^{(i)}}^\top(\by-\by^{(i)})
		\right).
	\end{equation}
	Rather than maximizing \(\widehat{J}_p\) directly, the method stabilizes the search around a current center \(\by^c\) by solving
	\begin{equation}\label{eq:bundle_master_root_cdc}
		\begin{aligned}
			\max_{\by,\eta}\quad
			&
			\eta-\frac{\rho}{2}\|\by-\by^c\|_2^2\\
			\text{subject to}\quad
			&
			\eta\le J(\by^{(i)})+{\bg^{(i)}}^\top(\by-\by^{(i)}),
			\qquad i=1,\dots,p,\\
			&
			\|\by\|_2\le 1,
		\end{aligned}
	\end{equation}
	where \(\rho>0\) is a proximal parameter. Problem \eqref{eq:bundle_master_root_cdc} is always feasible, for example \(\by=\by^c\) with \(\eta=\widehat{J}_p(\by^c)\), and eliminating \(\eta\) leaves a strongly concave maximization over a compact set, so its solution exists and is unique. The optimizer of~\eqref{eq:bundle_master_root_cdc} produces a new trial point. At that point, one evaluates \(J(\by)\) and the subgradient~\eqref{eq:subgradient} using exactly the same adjoint computation as in Approach II, appends the new cut to~\eqref{eq:bundle_model_root_cdc}, and repeats. The resulting estimate is
	\[
	\widehat{\delta}_{\rm bun}^{(K)}(\bx_T^\star)
	:=
	-\max_{0\le k\le K}J(\by^{(k)}).
	\]
	In practice, the bundle is truncated to at most \(p_{\max}\) cuts, while \(J\) and \(\bg\) are evaluated at each trial point with a composite trapezoidal rule on a uniform grid with \(N_q\) nodes. The ensuing case studies tests all of these methods. 
	
	\section{Numerical Experiments}
	\label{sec:casestudies}
	
	We apply the results of Lemma~\ref{lem:support} and Theorem~\ref{thm:margin} to linearized DAE models obtained from the IEEE 9, 39, 57, and 200 bus benchmarks under several renewable penetration and load levels. For each renewable penetration and load scaling pair, renewables enter as active power injections that offset part of the demand at the larger load buses; we solve the resulting nonlinear equilibrium, form the corresponding descriptor linearization \((\bE,\bA,\bB)\), set \(\bar{\bu}\) as a fraction of the equilibrium input magnitudes that shrinks with penetration and loading, generate the post-step initial condition from a small load increase and a simultaneous renewable drop, compute the corresponding free terminal state \(\bx_T^{\rm fr}\) over \(T=10~\mathrm{s}\), and then construct two DAE-consistent full state targets along a weak direction \(\by_{\rm cert}\) obtained from a Gramian or PBH analysis of \((\bA_r,\bB_r)\), placed at \(0.5\) and \(1.25\) times the directional support value. The latter is unreachable by construction since the dual objective in \eqref{eq:dual_distance_formula} is positive at \(\by_{\rm cert}\), and all screening studies below use these certified unreachable targets. An implementation of the three approaches, together with the linearized benchmark models, is available at \cite{code}. Tab.~\ref{tab:method_params} lists the algorithmic parameters.
	
	\begin{figure*}[t]
		\centering
		\includegraphics[width=0.99\textwidth]{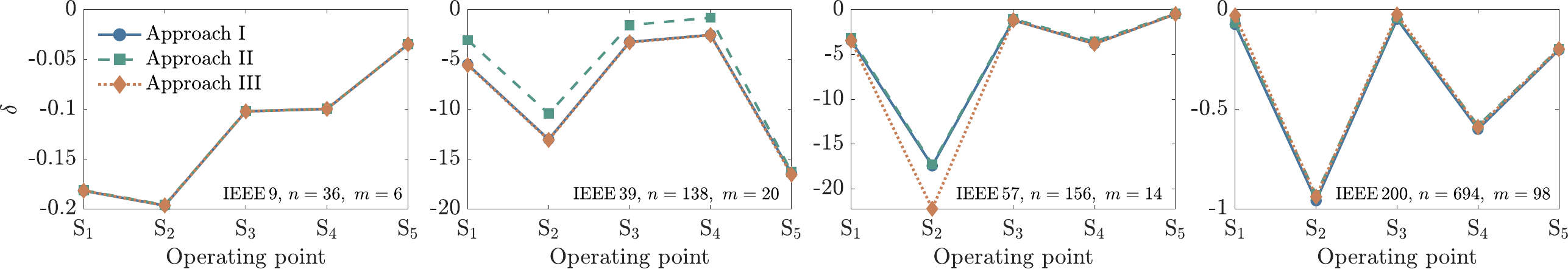}
		\caption{Comparison of the screening margin $\delta$ obtained with Approaches~I--III. From left to right: case9, case39, case57, case200. The x-axis indexes 5 scenarios shared across all cases where $S1$--$S5$ correspond to $(\mathrm{RenMean},\mathrm{LoadScale}) \in \{(0.05,1.00), (0.05,1.10), (0.25,1.00), (0.25,1.10), (0.50,1.10)\}$.}\label{fig:method_comparison}
		\vspace{-0.15cm}
	\end{figure*}
	
	\begin{table}[t]
		\caption{Case-study setup for the operating-point-dependent linearized DAE models. Common settings: \(\texttt{RenMean}\in\{0.05,\,0.25,\,0.50\}\), two terminal targets per operating point, and horizon \(T=10~\mathrm{s}\).}
		\label{tab:case_study_setup}
		\centering
		\scriptsize
		\setlength{\tabcolsep}{3pt}
		\renewcommand{\arraystretch}{1.1}
		\begin{tabular}{p{1.1cm}p{0.95cm}p{1.45cm}p{0.35cm}p{1.45cm}}
			\blueRule
			\rowcolor{mainblue!10}
			\textbf{Case} & \textbf{\((n,m)\)} & \makecell{\textbf{LoadScale}} & {\textbf{Ops.}} & \makecell{\textbf{\(\|\bar{\bu}\|_\infty\)}} \\
			\thinBlueRule
			IEEE 9 & \((36,6)\) & \makecell{1.00, 1.10,1.20} & 9 & \makecell{0.040 to 0.202} \\
			\thinBlueRule
			IEEE 39 & \((138,20)\) & \makecell{1.00, 1.05,1.10} & 7 & \makecell{0.865 to 4.274} \\
			\thinBlueRule
			IEEE 57 & \((156,14)\) & \makecell{1.00, 1.05,1.10} & 9 & \makecell{0.090 to 0.564} \\
			\thinBlueRule
			IEEE 200 & \((694,98)\) & \makecell{1.00, 1.05,1.10} & 9 & \makecell{0.077 to 0.510} \\
			\blueRule
		\end{tabular}
		\vspace{-0.25cm}
	\end{table}
	
	\begin{table}[t]
		\caption{Algorithmic parameters used in the numerical experiments. Symbols \(N_q\), \(K\), \(\alpha_k\), and \(\rho\) follow Section~IV; here \(\alpha_k=\alpha_0/\sqrt{k+1}\) and \(p_{\max}\) denotes the bundle cap.}
		\label{tab:method_params}
		\centering
		\scriptsize
		\renewcommand{\arraystretch}{1.1}
		\setlength{\tabcolsep}{3pt}
		\begin{tabular}{p{0.7cm}p{7.5cm}}
			\blueRule
			\rowcolor{mainblue!10}
			\textbf{Appr.} & \textbf{Settings} \\
			\thinBlueRule
			I & \(N_q=40\); SOCP solver tolerance \(10^{-8}\). \\
			\thinBlueRule
			II & \(N_q=80\), \(K=300\), \(\alpha_k=\alpha_0/\sqrt{k+1}\) with \(\alpha_0=1.0\), backtracking factor \(\beta=0.5\), backtracking cap \(8\), improvement tolerance \(10^{-10}\), normalized subgradient, and stall restart. \\
			\thinBlueRule
			III & \(N_q=80\), \(K=120\), proximal parameter \(\rho=2.0\), bundle cap \(p_{\max}=15\), and optimization tolerance \(10^{-8}\). \\
			\blueRule
		\end{tabular}
	\end{table}

	\subsection{Method Comparison}
	\label{subsec:method_comparison}
	We begin by comparing the three methods developed in Section~\ref{sec:compute} across all benchmark systems. For each operating point listed in Tab.~\ref{tab:case_study_setup}, the same full state terminal request \(\bx_T^\star\) is screened using Approaches~I--III and the resulting margins are recorded together with the corresponding computational times.
	
	Fig.~\ref{fig:method_comparison} compares the margins obtained from the three computational approaches across all benchmark systems and operating points. Approach~I serves as the conic benchmark because it directly solves~\eqref{eq:quad_margin} through the resulting SOCP~\eqref{eq:quad_socp}. As expected, margins obtained with Approaches~II and III remain close to those of Approach~I across all systems, indicating that the nonsmooth dual optimization in~\eqref{eq:J_of_y} captures the same separating geometry described in Theorem~\ref{thm:margin}. This confirms that the directional shortfall for power systems identified by the dual variable \(\by^\star\) is robust with respect to the numerical approach used to compute it. The underlying limitation is therefore structural (induced by actuator placement, network topology, and algebraic coupling) rather than some artifact of numerical discretization.
	
	\subsection{Convergence and Scalability}
	
	Tab.~\ref{tab:benchmark_relative_summary} summarizes the benchmark-relative convergence and computational scalability of the three screening methods. Approach~I is treated as the benchmark reference because for a sufficiently accurate quadrature rule it solves the conic approximation of the screening problem to high accuracy. Approaches~II and III are interpreted as first-order approximations to this benchmark rather than competing exact solvers.
	For each sampled infeasible instance, the certificate produced by Approach~I is denoted by $\delta^{\mathrm{I}}$. Approaches~II and III are declared to have reached the benchmark once their best-so-far certificate enters the tolerance band
	\[
	\bigl|\delta_{\mathrm{best}}^{(j)}(k)-\delta^{\mathrm{I}}\bigr|
	\le
	\varepsilon_{\mathrm{abs}}+\varepsilon_{\mathrm{rel}}\max\!\bigl(1,|\delta^{\mathrm{I}}|\bigr),
	\qquad j\in\{\mathrm{II},\mathrm{III}\},
	\]
	with $\varepsilon_{\mathrm{abs}}=10^{-5}$ and $\varepsilon_{\mathrm{rel}}=10^{-2}$. This is appropriate here because the objective is not exact iterate matching but practical agreement with the benchmark certificate at a scale commensurate with the certificate magnitude. The reported hit rate is the fraction of sampled instances that enter this band before the prescribed iteration cap.
	
	The results show a clear separation between the two first-order schemes. Approach~III reaches the benchmark tolerance band in very few iterations on the small and medium systems: the median count is $3$ iterations for IEEE~9 and $5$ iterations for IEEE~57, with $100\%$ hit rate in both cases. Even on IEEE~39, Approach~III reaches the benchmark within $15$ iterations at $100\%$ hit rate. The corresponding wall clock times remain below the benchmark runtime in all three systems. Approach~II is also faster than Approach~I in wall clock time, but its convergence is markedly less reliable under the same caps, hit rate drops to $20\%$ on IEEE~39 and IEEE~57, and the median benchmark-reaching iteration count is much larger when convergence occurs.
	
	For the large system (IEEE~200), Approach~III attains the benchmark band in median $5$ iterations and achieves a speedup exceeding one order of magnitude in wall clock time although its hit rate drops to $40\%$ under the fixed cap $K_{\mathrm{III}}=120$. Approach~II attains a higher hit rate of $80\%$ on this case but requires substantially more iterations and remains slower than Approach~III. This indicates that the bundle-type updates in Approach~III are substantially more effective at extracting a high-quality screening certificate early whereas the projected subgradient method can require many more iterations to achieve the same benchmark-relative accuracy.
	
	Overall, these results support the simple intended message of the paper. Approach~I provides a benchmark-quality certificate, while Approaches~II and III provide scalable approximations whose practical value is determined by how rapidly they approach that benchmark. Among the two first-order methods, Approach~III gives the strongest tradeoff between runtime and benchmark-relative accuracy, especially when only a small iteration budget is available.
	\begin{table}[t]
		\caption{Benchmark convergence and scalability across IEEE systems. Column I gives the average runtime of Approach~I. For Approaches~II--III, table reports average time to enter Approach-I tolerance band and the median iteration count to do so, hit rate in parentheses. Results averaged over 5 representative infeasible instances per system, excluding near-zero benchmark certificates. Iteration caps: $K_{\mathrm{II}}=300$ and $K_{\mathrm{III}}=120$.}\label{tab:benchmark_relative_summary}
		\centering
		\scriptsize
		\renewcommand{\arraystretch}{1.1}
		\setlength{\tabcolsep}{3pt}
		\resizebox{\columnwidth}{!}{
			\begin{tabular}{p{1.05cm}p{0.85cm}p{1.55cm}p{1.35cm}p{1.55cm}p{1.35cm}}
				\blueRule
				\rowcolor{mainblue!10}
				\textbf{Case} & \textbf{I [s]} & \makecell{\textbf{II to I [s]}\\\textbf{(speedup)}} & \makecell{\textbf{II med. iters}\\\textbf{(hit rate)}} & \makecell{\textbf{III to I [s]}\\\textbf{(speedup)}} & \makecell{\textbf{III med. iters}\\\textbf{(hit rate)}} \\
				\thinBlueRule
				IEEE 9 & 0.757 & 0.005\,(156.10$\times$) & 6.0\,(100\%) & 0.136\,(5.58$\times$) & 3.0\,(100\%) \\
				\thinBlueRule
				IEEE 39 & 0.760 & 0.221\,(3.44$\times$) & 100.0\,(20\%) & 0.315\,(2.41$\times$) & 15.0\,(100\%) \\
				\thinBlueRule
				IEEE 57 & 0.622 & 0.359\,(1.73$\times$) & 273.0\,(20\%) & 0.103\,(6.05$\times$) & 5.0\,(100\%) \\
				\thinBlueRule
				IEEE 200 & 18.172 & 3.456\,(5.26$\times$) & 123.0\,(80\%) & 0.329\,(55.19$\times$) & 5.0\,(40\%) \\
				\blueRule
		\end{tabular}}
		\vspace{-0.2cm}
	\end{table}

	\begin{figure}[t]
		\centering
		\includegraphics[scale=0.4]{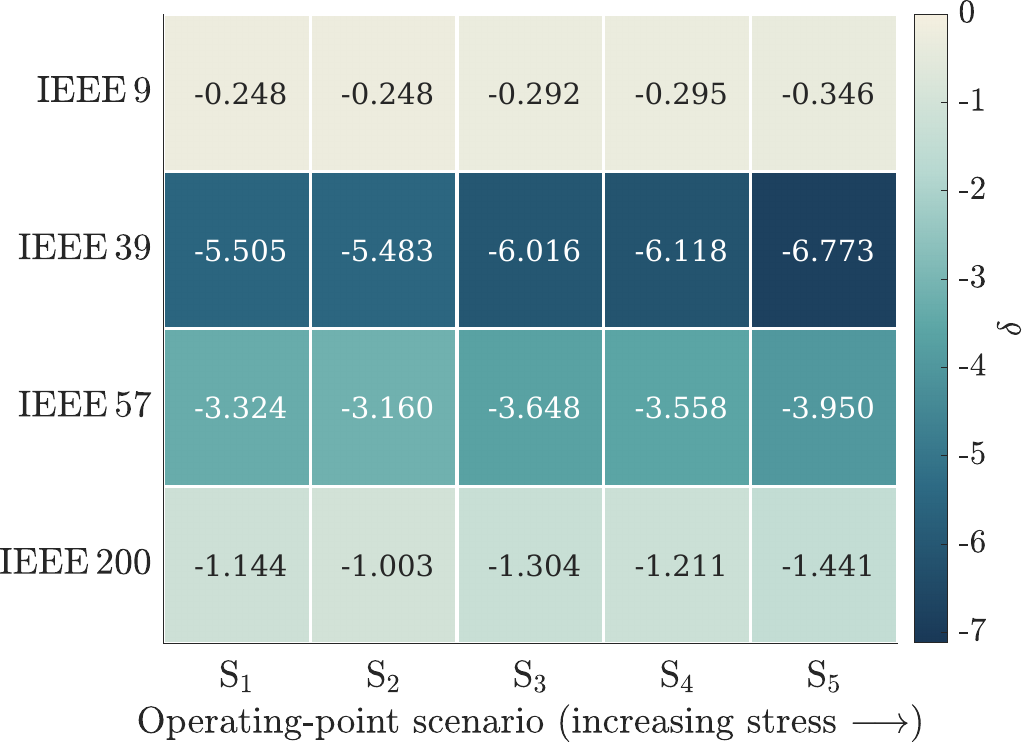}
		\caption{Effect of operating-point stress on the infeasibility margin \(\delta\). Increased renewable penetration and loading produce more negative margins, indicating reduced reachable terminal capability.}
		\label{fig:operating_stress}
		\vspace{-0.2cm}
	\end{figure}
	\subsection{Effect of Operating-Point Stress on Reachability}
	\label{subsec:stress_effect}
	We now study how operating-point stress affects the reachability margin. Stress is induced through simultaneous changes in renewable penetration and loading level, which modify both the Jacobian matrices and the effective control directions.
	
	Fig.~\ref{fig:operating_stress} shows the change of the screening margin as the operating point becomes more stressed. Across all tested systems, the margin becomes increasingly negative as renewable penetration and loading increase. This is consistent with a shrinking reachable set along the tested directions.
	
	From a control-theoretic standpoint, stressed operating points modify the reduced dynamics matrix \(\bA_r\) and reduce the effectiveness of the control matrix \(\bB_r\) along critical dynamic directions. In power system terms, increased renewable penetration replaces synchronous inertia with inverter-based resources which weakens electromechanical coupling and reduces controllability of frequency-related modes. 
	
	These combined effects reduce the directional support function in~\eqref{eq:full_support_formula} which leads directly to more negative values of \(\delta\). The resulting trend hence provides a quantitative interpretation of reduced controllability in stressed grids, not all terminal frequency or voltage recovery targets remain reachable within the prescribed time horizon and control envelope.
	
	\subsection{Influence of Control Limits and Terminal Demand}
	\label{subsec:control_limits}
	
	We now investigate how tightening the control envelope or increasing the terminal demand affects the reachability margin. The terminal request is parameterized along a reference direction $\bx_T(\alpha)=\bx_T^{\mathrm{fr}}+\alpha\,\boldsymbol{\xi}_T^{\mathrm{ref}}$ (where $\boldsymbol{\xi}_T^{\mathrm{ref}}$ is the displacement direction)
	so that increasing \(\alpha\) corresponds to requesting larger forced deviations from the free terminal response.
	
	\begin{figure}[t]
		\centering
		\includegraphics[width=0.99\columnwidth]{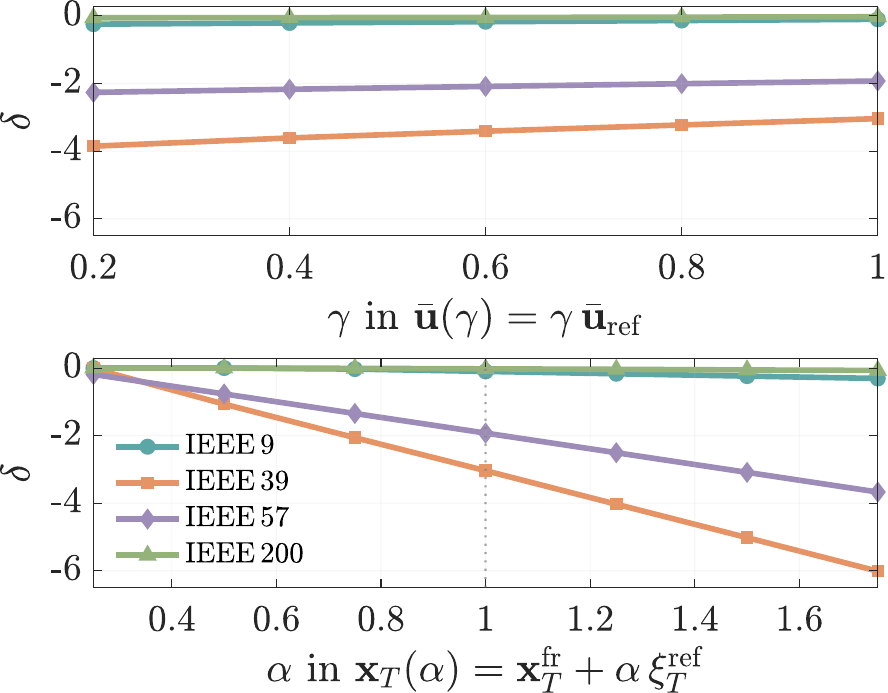}
		\caption{Effect of terminal demand scaling on the margin \(\delta\). Increasing the terminal request magnitude produces progressively more negative margins, indicating a growing control shortfall.}
		\label{fig:terminal_scaling}
		\vspace{-0.3cm}
	\end{figure}
	
	Fig.~\ref{fig:terminal_scaling} shows the margin obtained as the terminal request magnitude increases. For all systems, the margin decreases as \(\alpha\) grows. This matches the geometric interpretation of Theorem~\ref{thm:margin} as the requested terminal displacement moves farther from the reachable set, the Euclidean shortfall grows monotonically.
	
	A similar trend is observed when the control limits \(\bar{\bu}\) are reduced. Tightening the control envelope shrinks the reachable set through the scaling matrix \(\bR_{\bar u}\) in~\eqref{eq:full_support_formula}. This entails smaller control limits correspond to reduced governor authority, limited inverter headroom, or curtailed demand-response participation. Each of these conditions reduces the ability of the system to counteract disturbances within the finite horizon.
	
	These observations show that the screening margin captures both actuator limitations and target severity in a unified quantity. This enables operators to assess whether proposed frequency or voltage recovery objectives remain achievable under realistic actuator constraints without performing repeated time-domain simulations.
	
	\subsection{Comparison with CORA}
	\label{subsec:cora_comparison}
	We compare the proposed screening method with CORA \cite{CORA}. Since CORA natively propagates reachable sets for continuous dynamic models, we perform the comparison on the same reduced index-1 model used in \eqref{eq:alg_elim}--\eqref{eq:reduced_ode}, with the same initial state \(\bx_{d,0}\), horizon \(T\), and dynamic terminal target \(\bx_{d,T}^\star\). In CORA, the input box is represented as a zonotope and the terminal reachable set is computed as an outer approximation \(\widehat{\mathcal R}_{d,T}\) of the reduced dynamics. When the returned final set is represented as a zonotope, we denote its center by \(\bc_T\) and its generator matrix by \(\boldsymbol{\Gamma}_T\), so that
	\[
	\widehat{\mathcal R}_{d,T}
	=
	\left\{
	\bc_T+\boldsymbol{\Gamma}_T\mathbf{z}
	\;\middle|\;
	\|\mathbf{z}\|_\infty\le 1
	\right\}.
	\]
	Its support function is
	$
	h_{\widehat{\mathcal R}_{d,T}}(\by)=\by^\top \bc_T+\|\boldsymbol{\Gamma}_T^\top \by\|_1.
	$
	We then wrap the CORA output into the same target-specific margin geometry used in this paper by defining the outer set margin
	\[
	\delta_{\mathrm{CORA}}^{\mathrm{out}}(\bx_{d,T}^\star):=
	\min_{\|\by\|_2\le 1}
	\left(
	h_{\widehat{\mathcal R}_{d,T}}(\by)-\by^\top \bx_{d,T}^\star
	\right).
	\]
	If \(\delta_{\mathrm{CORA}}^{\mathrm{out}}<0\), then \(\bx_{d,T}^\star\notin \widehat{\mathcal R}_{d,T}\) and the reduced dynamic target is certified unreachable; if \(\delta_{\mathrm{CORA}}^{\mathrm{out}}\ge 0\), the result is inconclusive because \(\widehat{\mathcal R}_{d,T}\) is an outer approximation. The corresponding algebraic variables may then be recovered afterward through the linearized constraint map \(\bx_a=\bC_r\bx_d+\bD_r\bu_T\). Results of this comparison are reported in Tab.~\ref{tab:cora_comparison}.
	\begin{table}[t]
		\caption{Comparison between proposed approaches and a CORA-based outer set margin. Table reports average wall clock time over the sampled infeasible instances for each executed case and the corresponding average margin values.}
		\label{tab:cora_comparison}
		\centering
		\scriptsize
		\renewcommand{\arraystretch}{1.1}
		\setlength{\tabcolsep}{3pt}
		\resizebox{\columnwidth}{!}{
			\begin{tabular}{C{1cm}C{0.75cm}C{0.75cm}C{1.00cm}C{1.07cm}C{1.07cm}C{1.45cm}}
				\blueRule
				\rowcolor{mainblue!10}
				\textbf{Case} & \textbf{I [s]} & \textbf{III [s]} & \textbf{CORA [s]} & \makecell{\textbf{mean $\delta^{\mathrm{I}}$}} & \makecell{\textbf{mean $\delta^{\mathrm{III}}$}} & \makecell{\textbf{mean $\delta_{\mathrm{CORA}}^{\mathrm{out}}$}} \\
				\thinBlueRule
				IEEE 9 & 0.841 & 1.427 & 13.476 & -0.0279 & -0.02789 & -0.02716 \\
				\thinBlueRule
				IEEE 39 & 0.658 & 2.024 & -- & -1.433 & -1.384 & -- \\
				\thinBlueRule
				IEEE 200 & 1.806 & 4.961 & -- & -0.0631 & -0.0584 & -- \\
				\blueRule
		\end{tabular}}
	\end{table}

    Since \(\widehat{\mathcal R}_{d,T}\) is an outer approximation of the exact terminal set of the same reduced model, \(\delta_{\mathrm{CORA}}^{\mathrm{out}}\ge\delta^{\mathrm{I}}\) up to solver tolerances, and in Tab.~\ref{tab:cora_comparison} this ordering holds with gaps of \(1.5\%\) to \(8.8\%\) of \(|\delta^{\mathrm{I}}|\) on IEEE 9. The cost of the explicit set representation grows quickly with dimension. With CORA's adaptive algorithm at error tolerance \(10^{-2}\), the returned IEEE 9 terminal sets carry \(6{,}865\) to \(34{,}057\) zonotope generators at \(n_d=12\), and the same computation failed on IEEE 39 and IEEE 200. This is expected since tightness and tractability depend strongly on the time step, the zonotope order, and the reduction settings. In that sense, while CORA is a powerful general purpose reachability toolbox, it is less suitable for repeated screening of realistic power system models than the present margin computation which avoids explicit reachable set construction altogether. 
        
	
	\section{Paper Summary and Future Work}
	\label{sec:conclusion}
	
	The paper develops a DAE reachability screening framework for power systems in which the central object is a margin that uses the operating point, the horizon \(T\), the control limits \(\bar{\bu}\), and the linearized descriptor model \((\bE,\bA,\bB)\) to assess how far a demanding terminal target is from feasibility. The resulting margin supports rapid comparison of stressed and renewable-rich operating conditions before heavier reachable set construction or nonlinear simulation is carried out. Future extensions include operational path constraints on \([0,T]\) and using the formulation to place PMUs, actuation nodes, or HVDC lines to enlarge the terminal reachable set under heavy renewables or stressed conditions.

	\appendices
	
	\section{Proof of Lemma~\ref{lem:support}}
	\label{app:proof_lem}
		Every element of \(\mathcal{R}_T^{\rm full}\) is generated by the terminal relation \(\bx(T)=\bx_T^{\rm fr}+\mathbf{\Phi}_T(\bu,\bu_T)\), so it necessarily satisfies the algebraic closure induced by some admissible endpoint value \(\bu_T\); hence \(\mathcal{R}_T^{\rm full}\subseteq \mathcal{M}_T\). By definition of the support function and of \(\mathbf{\Phi}_T(\bu,\bu_T)\), \vspace{-0.1cm}
		\begin{equation*}
			\begin{aligned}
				h_{\mathcal{R}_T^{\rm forc}}(\by)
				=\;&
				\sup_{\bu\in\mathcal{U}_T,\ |\bu_T|\le \bar{\bu}}
				\Biggl[
				\int_0^T
				\bu(\tau)^\top\bB_r^\top e^{\bA_r^\top(T-\tau)}\bP^\top\by\,d\tau\\
				&\qquad\qquad\qquad
				+\bu_T^\top\bD_r^\top\by_a
				\Biggr].
			\end{aligned}
		\end{equation*}
		Let $\bq(\tau):=\bB_r^\top e^{\bA_r^\top(T-\tau)}\bP^\top\by$ and $\br:=\bD_r^\top\by_a.$
		Because the admissible set is a product of independent componentwise box constraints in time and at the endpoint, the maximization separates into a pointwise maximization for \(\bu(\tau)\) and a finite-dimensional maximization for \(\bu_T\). For almost every \(\tau\in[0,T]\),
		\[
		\begin{aligned}
			\sup_{|\bu(\tau)|\le \bar{\bu}}\bu(\tau)^\top\bq(\tau)
			&=
			\sum_{i=1}^m \sup_{|u_i(\tau)|\le \bar u_i} u_i(\tau)q_i(\tau)\\
			&=
			\sum_{i=1}^m \bar u_i |q_i(\tau)|
			=
			\|\bR_{\bar u}\bq(\tau)\|_1,
		\end{aligned}
		\]
		with a maximizer given componentwise by \(\bu^\star(\tau)=\bR_{\bar u}\,\mathrm{sign}(\bq(\tau))\). Since \(\bq(\tau)\) is continuous on \([0,T]\), \(\bu^\star(\tau)\) is measurable and satisfies \(|\bu^\star(\tau)|\le\bar{\bu}\) almost everywhere, so \(\bu^\star\in\mathcal U_T\) and the supremum over \(\mathcal U_T\) equals the integral of these pointwise suprema. Likewise,
		\[
		\begin{aligned}
			\sup_{|\bu_T|\le \bar{\bu}} \bu_T^\top\br
			&=
			\sum_{i=1}^m \sup_{|u_{T,i}|\le \bar u_i} u_{T,i}r_i = \sum_{i=1}^m \bar u_i |r_i|
			=
			\|\bR_{\bar u}\br\|_1,
		\end{aligned}
		\]
		with maximizer \(\bu_T^\star=\bR_{\bar u}\,\mathrm{sign}(\br)\). Substituting these expressions yields
		\begin{equation*}
			h_{\mathcal{R}_T^{\rm forc}}(\by)
			=
			\int_0^T
			\left\|
			\bR_{\bar u}\,\bB_r^\top e^{\bA_r^\top(T-\tau)}\bP^\top\by
			\right\|_1\,d\tau +
			\left\|
			\bR_{\bar u}\,\bD_r^\top\by_a
			\right\|_1,
		\end{equation*}
		which is exactly~\eqref{eq:full_support_formula}. Finally, since \(\mathcal{R}_T^{\rm full}=\bx_T^{\rm fr}\oplus\mathcal{R}_T^{\rm forc}\),
		\begin{equation*}
			\begin{aligned}
				h_{\mathcal{R}_T^{\rm full}}(\by)
				&=
				\sup_{\bz\in\mathcal{R}_T^{\rm forc}}
				\by^\top(\bx_T^{\rm fr}+\bz)\\
				&=
				\by^\top\bx_T^{\rm fr}
				+
				\sup_{\bz\in\mathcal{R}_T^{\rm forc}}\by^\top\bz =
				\by^\top\bx_T^{\rm fr}
				+
				h_{\mathcal{R}_T^{\rm forc}}(\by),
			\end{aligned}
		\end{equation*}
		which proves~\eqref{eq:full_support_shifted}.
	
	\section{Proof of Theorem~\ref{thm:margin}}
	\label{app:proof_thm}
		Lemma~\ref{lem:support} gives the explicit expression for \(h_{\mathcal{R}_T^{\rm forc}}(\by)\), so~\eqref{eq:ct_mcu_def} is a computable optimization problem. The remaining argument is geometric. Since \(\mathcal{R}_T^{\rm forc}\) is compact and convex, the projection \(\bp^\star:=\Pi_{\mathcal{R}_T^{\rm forc}}(\boldsymbol{\xi}_T)\) exists. If \(\bx_T^\star\in\mathcal{R}_T^{\rm full}\), then \(\boldsymbol{\xi}_T\in\mathcal{R}_T^{\rm forc}\). Hence \(h_{\mathcal{R}_T^{\rm forc}}(\by)\ge \by^\top\boldsymbol{\xi}_T\) for every \(\by\), so \(\delta(\bx_T^\star)\ge 0\). Choosing \(\by=\mathbf{0}\) in~\eqref{eq:ct_mcu_def} gives \(\delta(\bx_T^\star)\le 0\), hence \(\delta(\bx_T^\star)=0\).
		
		If \(\bx_T^\star\notin\mathcal{R}_T^{\rm full}\), then \(\boldsymbol{\xi}_T\notin\mathcal{R}_T^{\rm forc}\). Define
		\[
		\by^\star:=\frac{\boldsymbol{\xi}_T-\bp^\star}{\|\boldsymbol{\xi}_T-\bp^\star\|_2}.
		\]
		By the projection theorem,
		\[
		{\by^\star}^\top\bz\le {\by^\star}^\top\bp^\star
		\qquad
		\forall \bz\in\mathcal{R}_T^{\rm forc},
		\]
		so \(h_{\mathcal{R}_T^{\rm forc}}(\by^\star)={\by^\star}^\top\bp^\star\). Therefore,
		\[
		{\by^\star}^\top\boldsymbol{\xi}_T-h_{\mathcal{R}_T^{\rm forc}}(\by^\star)
		=
		{\by^\star}^\top(\boldsymbol{\xi}_T-\bp^\star)
		=
		\|\boldsymbol{\xi}_T-\bp^\star\|_2.
		\]
		For any \(\|\by\|_2\le 1\),
		\[
		\by^\top\boldsymbol{\xi}_T-h_{\mathcal{R}_T^{\rm forc}}(\by)
		\le
		\by^\top(\boldsymbol{\xi}_T-\bp^\star)
		\le
		\|\boldsymbol{\xi}_T-\bp^\star\|_2.
		\]
		Hence
		\[
		\max_{\|\by\|_2\le 1}
		\left(
		\by^\top\boldsymbol{\xi}_T-h_{\mathcal{R}_T^{\rm forc}}(\by)
		\right)
		=
		\mathrm{dist}\big(\boldsymbol{\xi}_T,\mathcal{R}_T^{\rm forc}\big).
		\]
		Rewriting this equality yields the second line of~\eqref{eq:screening_identity}. Finally, since \(\mathcal{R}_T^{\rm full}=\bx_T^{\rm fr}\oplus\mathcal{R}_T^{\rm forc}\) and \(\boldsymbol{\xi}_T=\bx_T^\star-\bx_T^{\rm fr}\), translation invariance gives
		\[
		\mathrm{dist}\big(\bx_T^\star,\mathcal{R}_T^{\rm full}\big)
		=
		\mathrm{dist}\big(\boldsymbol{\xi}_T,\mathcal{R}_T^{\rm forc}\big),
		\]
		which completes the proof.
	
	\section{Proof of Corollary~\ref{cor:monotonicity}}
	\label{app:proof_cor}
		If \(\bar{\bx}_T\in\mathcal{R}_{T,1}^{\rm full}\), then \(\bar{\bx}_T\in\mathcal{R}_{T,2}^{\rm full}\) by inclusion, and Theorem~\ref{thm:margin} gives \(\delta_1(\bar{\bx}_T)=\delta_2(\bar{\bx}_T)=0\). If \(\bar{\bx}_T\in\mathcal{R}_{T,2}^{\rm full}\setminus\mathcal{R}_{T,1}^{\rm full}\), then \(\delta_2(\bar{\bx}_T)=0\) while \(\delta_1(\bar{\bx}_T)<0\), so the claim again holds. If \(\bar{\bx}_T\notin\mathcal{R}_{T,2}^{\rm full}\), then \(\bar{\bx}_T\) lies outside both sets. Since distance to a set decreases when the set enlarges,
		\[
		\mathrm{dist}\big(\bar{\bx}_T,\mathcal{R}_{T,1}^{\rm full}\big)\ge
		\mathrm{dist}\big(\bar{\bx}_T,\mathcal{R}_{T,2}^{\rm full}\big).
		\]
		Applying Theorem~\ref{thm:margin} to both instances yields
		\[
		\delta_1(\bar{\bx}_T)
		=
		-\mathrm{dist}\big(\bar{\bx}_T,\mathcal{R}_{T,1}^{\rm full}\big)
		\le
		-\mathrm{dist}\big(\bar{\bx}_T,\mathcal{R}_{T,2}^{\rm full}\big)
		=
		\delta_2(\bar{\bx}_T).
		\]
		The magnitude statement follows immediately.
	
	\bibliographystyle{IEEEtran}
	\bibliography{refs}
	
\end{document}